\documentclass[aps,showpacs,nofootinbib,floatfix]{revtex4}

\usepackage{graphicx}

\usepackage{amsmath, amssymb, bm, graphicx, graphics, color,mathrsfs,multirow,ulem}

\newcommand{\be}{\begin{equation}}
\newcommand{\ee}{\end{equation}}
\newcommand{\ben}{\begin{eqnarray}}
\newcommand{\een}{\end{eqnarray}}

\newcommand{\cL}{{\cal L}}

\newcommand{\na}{\nabla}

\newcommand{\tpe}{{\tilde p}}

\newcommand{\hI}{\hat I}
\newcommand{\hg}{\hat g}
\newcommand{\hR}{\hat R}
\newcommand{\hna}{\hat \nabla}

\newcommand{\zpsi}{\psi^{\ast}}

\pacs{04.25.dg, 04.40.-b}
\begin{document}

\title{Dilatons and the Dynamical Collapse of Charged Scalar Field}


\author{Anna Nakonieczna and Marek Rogatko}
\affiliation{Institute of Physics \protect \\
Maria Curie-Sklodowska University \protect \\
20-031 Lublin, pl.~Marii Curie-Sklodowskiej 1, Poland \protect \\
aborkow@kft.umcs.lublin.pl \protect \\
rogat@kft.umcs.lublin.pl \protect \\
marek.rogatko@poczta.umcs.lublin.pl}

\date{\today}

\begin{abstract}
We studied the influence of dilaton field on the dynamical collapse of a charged scalar one.
Different values of the initial amplitude of dilaton field as well as the altered
values of the dilatonic coupling constant were considered.
We described structures of spacetimes and properties of black holes emerging from the collapse of 
electrically charged scalar field in dilaton gravity. Moreover, we provided a meaningful comparison of the 
collapse in question with the one in Einstein gravity, when dilaton field is absent and 
its coupling with the scalar field is equal to zero.
The course and results of the dynamical collapse process seem 
to be very sensitive to the amplitude of dilaton field and to the value
of the coupling constant in the underlying theory.
\end{abstract}

\maketitle

\section{Introduction}
\label{intro}
Gravitational collapse and black hole {\it no-hair}
conjecture or its mathematical formulation black hole uniqueness theorem
attracted attention for many decades of researches. The uniqueness theorem
states that stationary axisymmetric solution of Einstein-Maxwell equations relaxes to Kerr-Newman 
spacetime and is characterized by black hole mass, charge and angular momentum \cite{uniq}.
The extension of the aforementioned problems to n-dimensional spacetime was also achieved \cite{ndim}.
Unfortunately, this simple picture is in contrast with the black hole
interior description. The singularity theorem \cite{haw73} predicts occurrence of spacetime 
singularities as the result of a gravitational collapse, while the weak cosmic censorship conjecture
\cite{pen} states that
these kind of singularities are hidden below the black hole event horizon.
\par
Quite different picture of the inner black hole singularity emerged \cite{his81}-\cite{is} last years. 
It turns out that the Cauchy
horizon inside charged or rotating black hole is transformed into null weak singularity
in the sense that an infalling observer who hits this null singularity experiences only 
a finite tidal deformation \cite{ori,bra95}.
One also encounters {\it mass inflation} phenomenon, when the curvature scalars diverge at the Cauchy horizon.
The physical mechanism on which {\it mass inflation}
is based, is bounded with small perturbations being the 
remnants of gravitational collapse which are gravitationally blue-shifted as they propagate in the black hole
interior parallel to the Cauchy horizon \cite{is}. The above conclusions were
mainly based on the perturbative analysis.
\par
On the other hand,
the first step in the direction of studies of a full nonlinear investigations of the inner structure
of black holes was the research of Gnedin and Gnedin \cite{gne93}, where they showed the existence of 
a central singularity inside a charged black hole. Hamade and Stewart \cite{ham96} studied numerically
the spherically symmetric collapse of massless field. 
They took into account the initial data depending on a parameter. It was
shown that the field in question either dispersed to infinity or collapsed
to a black hole, depending on the strength of the initial data.
Brady and Smith \cite{bra95} provide the studies of the nonlinear evolution of the neutral 
scalar field on a spacetime of charged black hole. In \cite{bur98} the divergence rate of the
blue-shifted factors, valid along the Cauchy horizon, were elaborated analytically. 
However, the aforementioned numerical studies began on 
Reissner-Nordstr\"om (RN) spacetime and the black hole formation and the 
{\it mass inflation}  were not shown explicitly.
\par
Numerical studies of the massless scalar field in the case of spherically symmetric spacetime
were elaborated in \cite{aya97}. On the other hand, in \cite{hod98} it was explicitly
demonstrated the the {\it mass inflation} occurred during a dynamical charged gravitational collapse.
Starting with the regular spacetime, the evolution through the formation of an apparent horizon, 
then the Cauchy horizon and a final singularity was performed. In \cite{ore03}  the evolution of a collapsing spherical shell
of a charged massless scalar field was analyzed, and an external RN spacetime as well as an inner spacetime
bounded by a singularity on the Cauchy horizon was obtained. The aforementioned results were refined and confirmed
in \cite{han05}, where the physics of nonlinear processes inside the spherical charged black hole perturbed
by minimally coupled massless scalar field was studied.
\par
Pair creation process in the strong electric field in the dynamical collapse of a self-gravitating
electrically charged massless scalar field was treated numerically in \cite{sor01}. Moreover, the insight into
the dynamical formation and evaporation of a spherically charged black hole emitting Hawking radiation
was presented in \cite{sor01b}. The further studies along the line mentioned above were provided in \cite{hon10},
where a spherically symmetric charged black hole with a complex scalar field, gauge field and the 
normalized energy-momentum tensor were taken into account. Among all it was shown that the Hawking radiation
caused the inner horizon was separated from the Cauchy one. On the other hand,
the neutralization of the charged black hole forced the inner horizon to evolve directly into a spacelike
singularity, while the Cauchy horizon became a null singulariy tending towards the spacelike one. The 
behaviour of the Brans-Dicke field
during gravitational collapse of matter was analyzed in \cite{hwa10}. In order to consider
Hawking radiation in the spherically symmetric dynamical collapse the normalized
energy-momentum tensor was used in \cite{hwa11}. It enables to analyze the discharge and neutralization of a 
charged black hole both non evaporating and evaporating.
An alternative approach to the problem of gravitational collapse was proposed in \cite{racz06}. On account of the careful 
treatment of trapped regions it was possible to investigate the role of topology changes during the examined 
process \cite{racz10}.
The first axisymmetric numerical code testing the gravitational collapse of 
a complex scalar field was presented in \cite{sor10}.
It turns out that also the non-linear processes with the participation of an exotic scalar field 
modeled as a free scalar field
with an opposite sign in the energy-momentum tensor, were considered \cite{dor10} due to the case when
RN black hole was irradiated by this kind of matter.
Recently, numerical simulations of five-dimensional spherically symmetric gravitational collapse of massless
scalar field in Einstein-Gauss-Bonnet gravity were conducted in \cite{gol12}. \\
It turned out that studies of the dynamical surface gravity in general spherical setting enabled numerical studies of black hole formation due to a scalar field collapse \cite{pie11}. On the other hand, a Hamiltonian formulation of spherically symmetric scalar field collapse helped to include quantum correction in the aforementioned studies \cite{zip10}.\\
A detailed analytical discussion concerning the structures of spacetimes containing realistic black holes 
formed during a gravitational collapse was presented in \cite{daf12}. It surprisingly turned out that 
the null singularity along the Cauchy horizon may be not followed by a spacelike one.
\par
As far as the uniqueness theorem for the low-energy string black holes is concerned, various aspects of this problem
were treated in the case of four-dimensional spacetime \cite{len}. Moreover,
the late-time and intermediate behaviour of scalar and fermion fields in the background of dilaton and 
Euclidean dilaton black holes were investigated in \cite{fiel}.\\
The implication of the superstring theory on the dynamical process of the gravitational collapse was studied in
\cite{bor11}. Namely, the gravitational collapse of a self-interacting charged scalar field in the 
context of the low-energy string theory, the so-called dilaton gravity was considered. The numerical studies
revealed that there was no formation of the inner horizon. The collapse in question resembled the collapse 
leading to the formation of Schwarzschild black hole spacetime.
\par
In our paper we shall study the influence of the dilaton field amplitude 
on the dynamical collapse of the charged scalar field.
We choose the constant amplitude of the charged scalar field and conduct our studies 
for various values of the dilaton field amplitude. One also alters the value of 
the coupling constant in the theory under inspection. In what follows 
we assume that the considered Lagrangian for the charged complex scalar fields will be
coupled to the dilaton via an arbitrary coupling, i.e.,
$e^{2 \alpha \phi} \cL (\psi, \zpsi, A)$, in the {\it string frame}.\\ 
The outline of the remainder of the paper is as follows.
In Sec.II we describe the model we shall work on.
Sec.III will be devoted to the numerical scheme
applied in our investigations. We discussed the numerical algorithm, an adaptive grid used in computations and
the boundary and initial conditions for the equations of motion for the considered problem. Sec.IV 
is assigned to the discussion of the obtained results. In Sec.V we concluded our researches.

\section{Model of a dynamical black hole collapse}
Recently, there has been a revived interest in the exact solutions in coupled system with massless scalar
field called dilaton \cite{dil}. These studies were mainly motivated by the 
low-energy limit of the heterotic string theory, the so-called dilaton gravity. In this theory dilaton
field is coupled to the other fields in a non-trivial way. To account for the unknown
coupling constant one chooses the underlying action in the form as (see, e.g., \cite{gre97})
\be
\hI = \int d^{4} x \sqrt{- \hg} \left [ e^{- 2 \phi}
\left (
\hR - 2 \left ( \hna \phi \right )^2 + e^{ 2 \alpha \phi} \cL \right )
\right ],
\label{a1}
\ee
where $\phi$ is the dilaton field and $\alpha$ is the coupling constant.
Our interests will concentrate on the dynamical collapse of the charged complex scalar field when
gravitational interactions will constitute the dilaton gravity.
Therefore the gauge invariant Lagrangian for the complex scalar field 
$\psi$ coupled to the $U(1)$-gauge field
$A_\mu$ will be provided by
\be
\cL = - {1 \over 2} \left (
\hna_{\alpha} \psi + i e A_{\alpha} \psi \right ) \hg^{\alpha \beta}
\left (
\hna_{\beta} \zpsi - i e A_{\beta} \zpsi \right ) - F_{\mu \nu}F^{\mu \nu} .
\label{lag}
\ee
The above action is written in in the so-called {\it string frame}, the usual way it appears in the string
sigma model. In order to get the gravitational part of the action in a more familiar form,
one needs to rewrite it to the {\it Einstein frame} performing a conformal transformation
\be
g_{\alpha \beta} = e^{- 2 \phi} \hg_{\alpha \beta}.
\ee
After the transformation we may write the underlying action (\ref{a1}) in the form
\be
I = \int d^{4} x \sqrt{-g} \bigg[
R - 2 ( \na \phi )^{2} + e^{2 \alpha \phi + 4 \phi} \cL
(\psi, \zpsi, A, e^{2 \phi} g_{\alpha \beta}) \bigg].
\label{a2}
\ee
It turns out that the same form of the action arises from five-dimensional Kaluza-Klein theory \cite{kk}.
Namely, one can recast the five-dimensional metric in the canonical form provided by
\be
ds^2 = e^{-{4\phi \over \sqrt{3}}}~\bigg( dx^5 + 2~A_{\mu}~dx^\mu \bigg)^2 + 
e^{{4\phi \over \sqrt{3}}}~g_{\mu \nu}~dx^\mu~dx^\nu,
\ee
where $x^\mu$ are the four-dimensional coordinates. This decomposition into four-dimensional
is such that fields do not depend on the fifth dimension. The five-dimensional
Einstein-Hilbert action up to the surface terms can be written as
\be
S_{KK} = \int d^5 x \sqrt{- {}^{(5)}g}~R = \int d^{4} x \sqrt{-g} \bigg[
R - 2 ( \na \phi )^{2} - e^{- 2~\sqrt{3}~\phi}~F_{\mu \nu}F^{\mu \nu} \bigg].
\ee
Just the general form of the considered action (\ref{a2}) enables one to study dynamical collapse
of charged complex scalar field in various gravity theories. One has the dilaton gravity when $\alpha = - 1$. When
$\alpha = - \sqrt{3}$ the theory stems from five-dimensional Kaluza-Klein theory, while the case
$\alpha = 0$ is responsible for the standard Einstein-Maxwell theory.\\
In order to treat the problem in question we choose $(2+2)$-spherically symmetric double-null
line element \cite{chr93}
\be
ds^2 = - a(u, v)^2 du dv + r^2(u, v) d \Omega^2,
\label{m}
\ee
where $v, u$ are advanced and retarded time null coordinates.
Moreover, the choice of the coordinates enables us to follow the evolution from the region where the 
spacetime is regular (approximately null infinity), through the formation of horizons and further
to the final central singularity.
\par
The underlying equations of motion are derived from the variational principle. The consequence of the choice of the
double-null coordinates for the electromagnetic field and electromagnetic potential as well the detailed
description of the equations of motion were described in our previous paper \cite{bor11}.

\section{Numerical computations}
It turns out that 
an analytical solution of the set of conjugate equations of motion in the theory under consideration
is unobtainable.  Therefore numerical methods ought to be used in order to draw conclusions about the structure 
of spacetime emerging from the examined evolution.

In the present paper we have used the numerical algorithm containing adaptive mesh refinement.
The particulars about the method are 
described in Ref.\cite{bor11}. The manner of setting the boundary and initial conditions 
for considered equations remained unchanged in comparison to \cite{bor11}. The one-parameter families 
representing initial profiles of the field functions used for simulations considered in the present paper 
are provided by
\ben \label{profiles1}
f_D &=& \tpe~ v^2~ e^{-\left(\frac{v-c_1}{c_2}\right)^2},\\ \label{profiles2}
f_S &=& \tpe~ \sin^2 \left(\pi\frac{v}{v_f}\right)~
\Bigg( \cos\left(\pi\frac{2v}{v_f} \right) + i~\cos \left(\pi\frac{2v}{v_f}+\delta\right)\Bigg).
\een
The $f_D$-family refers to the dilaton 
field with the family constants $c_1=1.3$ and $c_2=0.21$. 
On the other hand, the $f_S$-family describes the 
electrically charged scalar field. The family constant $v_f=7.5$ and the parameter 
determining the amount of the initial electric charge $\delta=\frac{\pi}{2}$. 
In both cases a free family parameter is denoted by $\tpe$.\\
Unfortunately, there are no analytical solutions neither for the problem in question nor for any of simplified 
versions of it (apart from the extremely simplified, trivial case of empty spacetime). It implies that checking 
the correctness of the numerical code has to be based on indirect methods.
\par
The three tests used for proving the credibility of the code and justifications for carrying 
them out are widely discussed in \cite{bor11}. As was reported in the aforementioned paper, the results obtained 
for the different integration steps on the non-adaptive grid 
display the satisfying agreement of an order of $0.01\%$. Moreover, the code 
demonstrates the
linear convergence as is expected for the applied algorithm.
It also reveals the errors decrease with the increase of the grid density.
\par
The issue of mass and charge conservation was also analyzed. 
Because of the fact that the dynamical collapse of charged scalar field in dilaton gravity resembles
Schwarzschild one \cite{bor11}, we define the mass function as the mass included 
in a sphere of the radius $r\left(u,v\right)$. It yields
\be
m(u,v) = {r \over 2}~\bigg( 1 + {4 ~r_{,u}~r_{,v} \over a^2} \bigg), 
\label{mhaw}
\ee
while the electric charge is defined by the relation
\be
Q = {2~r^2 \over a^2}~A_{u,v}.
\ee
We obtained profiles of $m$ and $Q$ versus $u$ along ingoing null rays corresponding to the future null infinity, 
that is $v_f=7.5$ in our computations. The obtained profiles appeared to be in a qualitative agreement with those
presented in Ref.\cite{bor11}. For evolutions leading to the formation of a black hole, the mass was conserved 
within $1.6\%$ and the electric charge within $2.5\%$, apart from regions in the vicinities of the horizons. 
The issue of conservation laws in the nearby of the horizons was discussed in \cite{bor11}.
\par
\begin{figure}[p]
\includegraphics[scale=0.45]{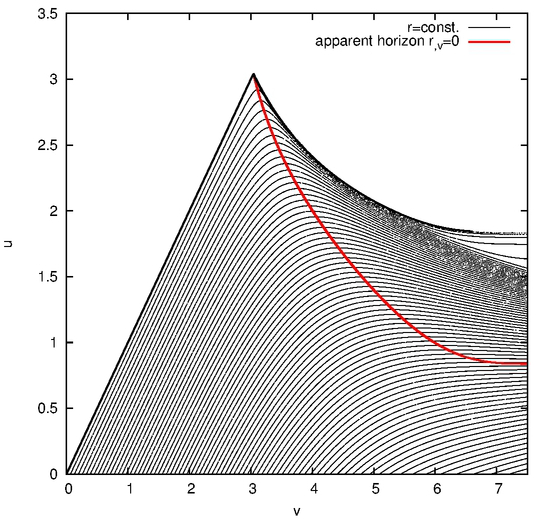}
\caption{Lines of constant $r$ in the $(vu)$-plane for the evolution of electrically charged scalar field. 
The family parameter is given by $\tpe_s=0.6$.}
\label{fig01}
\end{figure}

Further, we examined the simplified versions of the considered problem, i.e.,
the dynamical collapses of neutral and 
electrically charged scalar fields leading to Schwarzschild and RN spacetimes, respectively. 
The evolutions were described in detail in our previous paper. Our results are consistent with 
those obtained in Refs.\cite{ham96,hod98,ore03}. In what follows, the Penrose diagram of the 
latter spacetime shall be essential for our further analysis. It is depicted in Fig.\ref{fig01}.

\section{Results}
In our numerical computations we have used the one-parameter families of the field profiles
provided by relations (\ref{profiles1}) and (\ref{profiles2}). 
For the brevity of notation, the free family parameter $\tpe$ will be written with subscript referring to type of the 
considered field, i.e., $\tpe_s$ 
will relate to electrically charged scalar field while $\tpe_k$ will stand for dilaton field.
\par
As was stated in the introduction, the main aim of our research was to investigate the 
results of a collapse of an electrically charged scalar field in the presence of the dilaton field
for different dilaton field initial amplitudes $\tpe_k$ and different values of the
coupling constant $\alpha$. We will analyze the cases of $\alpha$ equal to $-\sqrt{3}$, $-1$, $0$ and $+1$. 
The examined range of the $\tpe_k$-values varies from $0$ to $0.13$. The upper limit is close to the maximum value, which 
allows us to begin our computations in the region outside the event horizon.
\par
As was mentioned while introducing the model under consideration, the exact value of $\alpha$ is unknown. However, 
our choice of the values of the dilatonic coupling constant is dictated by the correspondence 
between these particular values and analytical 
models discussed in the literature. Specifically, $\alpha=-\sqrt{3}$ refers to Kaluza-Klein theory, $\alpha=-1$ 
is connected with the low-energy string theory and when $\alpha=0$ the dynamical collapse of an electrically 
charged scalar field in presence of an uncoupled dilaton field is obtained. The last value taken into account, 
$\alpha=+1$, seems to be unphysical. However, we regard it as interesting to examine whether 
the results of a collapse with $\alpha=+1$ are dramatically different from these obtained for the other values.
\par
The presentation and interpretation of the results will be carried out in the three stages. At first, we shall
establish the reference point with which we will compare our further results. We will describe its spacetime 
structure (type, locations of the horizons and singularity origin) and introduce the features of an 
intrinsic black hole (radius, mass and charge) that will be essential in the following analysis. Afterwards, we 
will depict and describe spacetime structures obtained during evolutions for the considered values of $\alpha$ 
and compare them with the case of reference. Finally, we will comment on the features of black holes obtained 
in spacetimes representing different values of dilatonic coupling constant. One also compares 
the resulting black holes with those in the reference spacetime.

\subsection{Reference point}
To begin with, we shall establish our point of reference. Because of the fact that the amplitude of the dilaton 
field will
vary in different evolutions, we choose an evolution with $\tpe_k=0$ as our reference point.
This condition corresponds to the 
dynamical collapse of the only electrically charged scalar field.
The free family parameter characterizing the initial amplitude of the electrically charged scalar field 
is chosen to be equal to $\tpe_s=0.6$ and will be constant for all evolutions described in this paper. 
Moreover, we assigned the value of $0.5$ to the electric coupling constant $e$ in all considered evolutions.
One recalls that the results of the dynamical collapse do not depend on it \cite{bor11}. 
The structure of the reference spacetime 
is the dynamical RN spacetime.
It is presented in Fig.\ref{fig01}. In the Penrose diagram of constant $r\left(u,v\right)$ 
in the $\left(vu\right)$-plane we observe an apparent 
horizon in the domain of integration, which after a dynamical part of the evolution
settles along an event horizon for $v\rightarrow\infty$. 
The outermost line corresponds to $r=0$, whose singular part situated on the 
right side of the peak refers to spacelike singularity. A collection of lines 
$r=const.$ situated beyond the event horizon, each of which settles along constant $u$-coordinate indicates the existence of a Cauchy horizon at $v\rightarrow\infty$.\\
In the reference case, the line $r=0$ becomes 
singular at $u=3.04$ (the peak on the Penrose diagram of spacetime). We will refer to this point as 
singularity origin. The location of an event horizon corresponds to $u=0.84$. The radius of the 
considered black hole is the radius of the event horizon and is equal to $1.63$. Moreover, the two 
features characterizing RN spacetime, namely mass and charge of a black hole, are in our case 
equal to $0.94$ and $-0.64$, respectively. These are the values corresponding to the final, non-dynamic 
part of the considered evolution, i.e., they are calculated at the event horizon at $v\rightarrow\infty$. 
The Hawking mass for RN black hole implies
\be
m(u,v) = {r \over 2}~\bigg( 1 + {4 ~r_{,u}~r_{,v} \over a^2} + \frac{Q^2}{r^2} \bigg),
\label{mhaw-RN}
\ee
contrary to the mass 
described by the relation
(\ref{mhaw}) and corresponding to the Schwarzschild-type spacetimes.

\subsection{Structures of spacetimes}
\subsubsection{Negative dilatonic coupling constant}
\begin{figure}[p]
\includegraphics[scale=0.45]{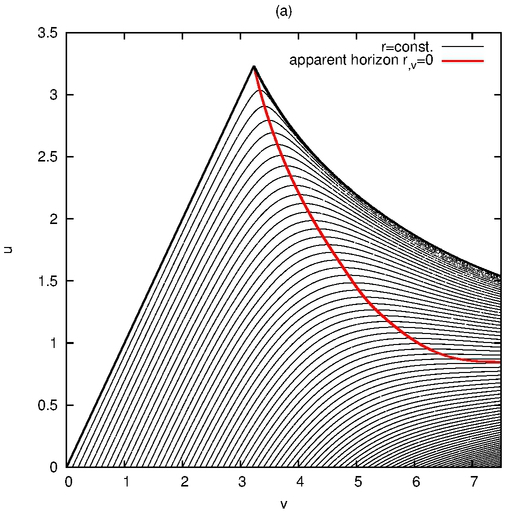}\includegraphics[scale=0.45]{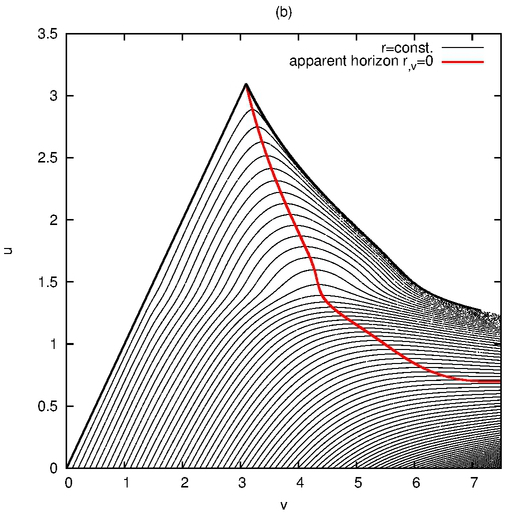}
\includegraphics[scale=0.45]{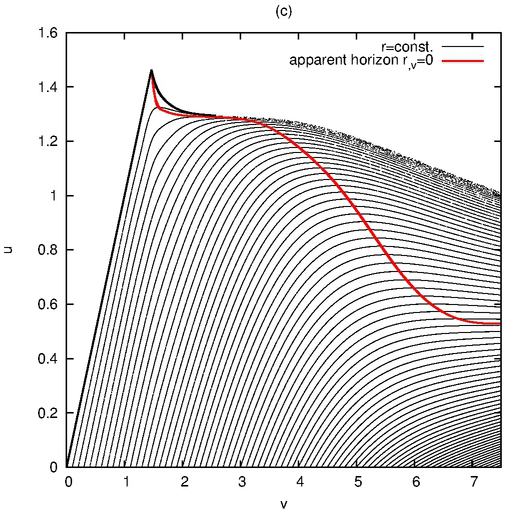}\includegraphics[scale=0.45]{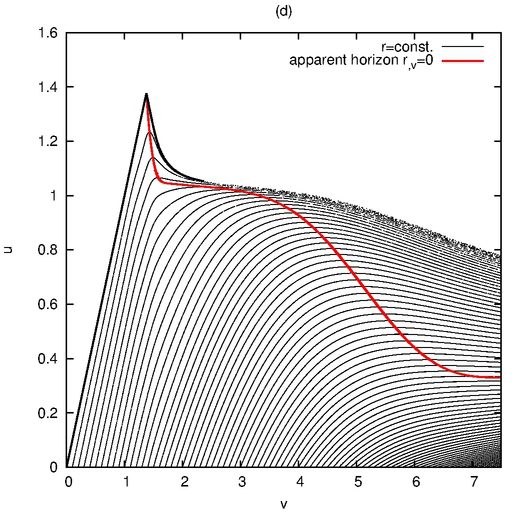}
\caption{Lines of constant $r$ in the $(vu)$-plane for the 
dynamical collapse of charged scalar field. The electric and the dilatonic coupling constants 
equal to $e=0.5$ and $\alpha=-\sqrt{3}$, respectively. Family parameter for the electrically charged scalar field 
is set $\tpe_s=0.6$, while for dilaton field are chosen as 
(a) $\tpe_k=0.01$, (b) $\tpe_k=0.05$, (c) $\tpe_k=0.075$ and (d) $\tpe_k=0.1$.}
\label{fig02}
\end{figure}

\begin{figure}[p]
\includegraphics[scale=0.45]{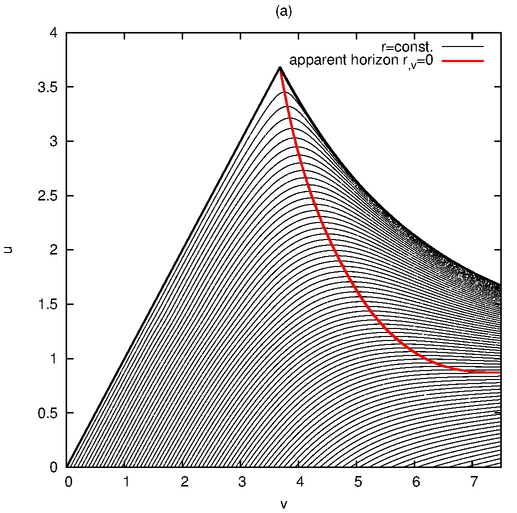}\includegraphics[scale=0.45]{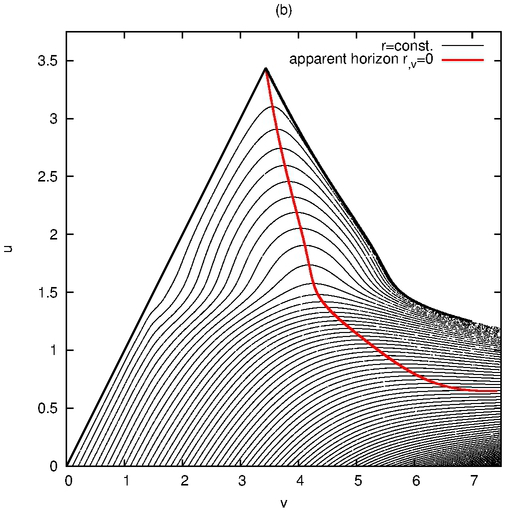}
\includegraphics[scale=0.45]{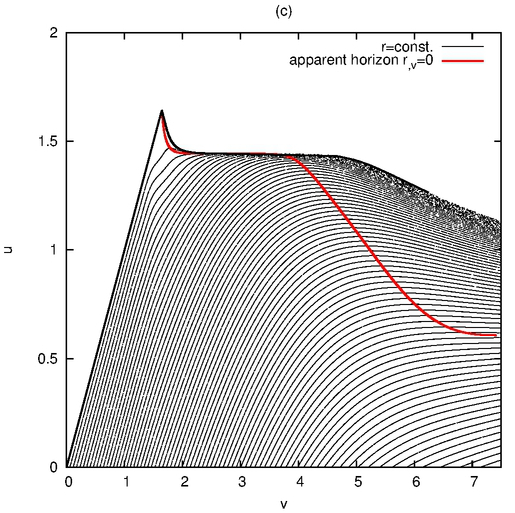}\includegraphics[scale=0.45]{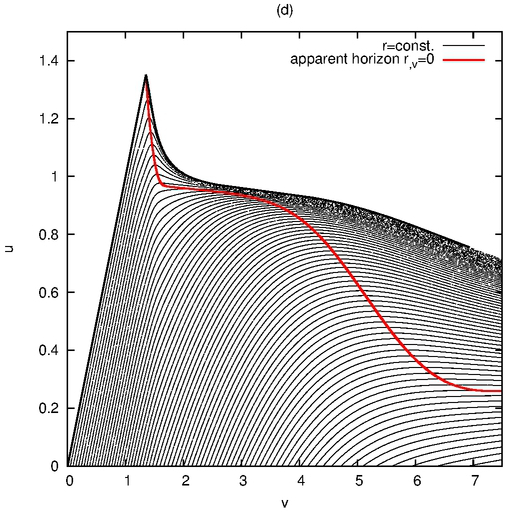}
\caption{Lines of constant $r$ in the $(vu)$-plane for the 
dynamical collapse of charged scalar field. The electric and the dilatonic coupling constants 
equal to $e=0.5$ and $\alpha=-1$, respectively. Family parameter for the electrically charged scalar field 
is set $\tpe_s=0.6$, while for dilaton field are chosen as 
(a) $\tpe_k=0.03$, (b) $\tpe_k=0.065$, (c) $\tpe_k=0.07$ and (d) $\tpe_k=0.11$.}
\label{fig03}
\end{figure}

The structures of spacetimes emerging from the collapse of an electrically charged scalar field coupled 
to the dilaton field for two considered negative values of dilatonic coupling constant, that is 
for $\alpha=-\sqrt{3}$ and $\alpha=-1$, are qualitatively similar, so they will be analyzed altogether. 
The Penrose
diagrams of spacetimes resulting from the considered collapse for $\alpha=-\sqrt{3}$ are shown in 
Fig.\ref{fig02}. The free family parameter for the dilaton field representing its initial 
amplitude was set as equal to $0.01$, $0.05$, $0.075$ and $0.1$. In Fig.\ref{fig03} we 
presented lines of constant $r$ in the $(vu)$-plane for the 
case of the $\alpha$ coupling constant equal to $-1$.
The initial amplitudes of the dilaton field $\tpe_k$ were chosen to be equal to 
$0.03$, $0.065$, $0.07$ and $0.11$. The most important feature of 
the dynamical collapse of the charged scalar field in the considered cases is the fact that
the presence of the dilaton field results in the disappearance of the Cauchy horizon 
of the newly formed black hole spacetime in comparison with the case of $\tpe_k=0$ (RN spacetime). Even for 
very small values of the parameter $\tpe_k$ 
the resulting spacetime has only one horizon surrounding the central spacelike singularity. 
Despite of the presence of an electric charge in the spacetime in question, 
its structure corresponds to the dynamical Schwarzschild-type spacetime.\\
In both cases of negative values of $\alpha$ coupling constant, for small values of the parameter $\tpe_k$, the
typical Schwarzschild-like spacetime is obtained. The apparent horizon smoothly changes its 
location towards smaller values of $u$-coordinate as the advanced time increases and it reaches 
the event horizon at $v\rightarrow\infty$. For larger values of the initial dilaton field 
amplitude a bend in the apparent horizon appears. Hence its course cease being so smooth. Such a situation 
is visible for $\tpe_k=0.05$ and $\tpe_k=0.065$ for $\alpha$ equal to $-\sqrt{3}$ and $-1$ 
(Figs.\ref{fig02}b and \ref{fig03}b), respectively. In both cases there is a 
tendency towards black hole formation at around $u=1.5$ (bends in lines $r=const.$ corresponding to 
the bend in the apparent horizon). But the strength of the dilaton field is still too small 
to influence drastically
the structure of spacetime and the line $r=0$ remains non-singular until 
significantly larger values of the retarded time, where the black hole eventually forms. The situation changes 
for the larger initial dilaton field amplitudes. After the dynamical part of the evolution (significant 
decrease of $u$-coordinate along an apparent horizon) the horizon temporarily settles along a constant 
value of $u$ and then the second dynamical stage of the collapse takes place. During this phase the horizon 
changes its position to another $u=const.$, which is in fact smaller retarded time. 
We refer to the part of an apparent horizon, which settles along constant $u$ between the two 
dynamical stages of a collapse as a temporary horizon.
As may be inferred from 
Figs.\ref{fig02} 
and \ref{fig03} the appearance of the temporary horizon is accompanied by the fact that the line $r=0$ 
becomes singular significantly earlier in terms of the retarded time $u$.
\par
In order to thoroughly examine the influence of the initial dilaton field amplitude on locations of 
the event horizon, temporary horizon and singularity origin we conducted computations for a 
wide range of values of $\tpe_k$. The results are shown in Fig.\ref{fig06}, where plots a and b 
refer to $\alpha=-\sqrt{3}$ and $\alpha=-1$, respectively.\\ 
We shall now make a short comment on the manner of determining the considered $u$-locations. The values 
of retarded time characterizing the event horizon and singularity origin are simply the values of $u$ 
corresponding to the apparent horizon at $v=v_f$ and the peak of the Penrose diagram, respectively. The 
location of a temporary horizon is less obvious to determine. We based the process on examining the changes 
of $u$ and $v$-coordinates along the apparent horizon. The changes of retarded time ($\delta u$) are constant 
due to the construction of the numerical grid. The changes of advanced time ($\delta v$) vary from $\delta u$ 
to the values of an order of $10^{3}\delta u$. The values of $\delta v$ close to $\delta u$ indicate dynamical 
parts of the evolution, while the growth of $\delta v$ means that the apparent horizon settles along $u=const.$ 
and the non-dynamical part of evolution appears. Such big changes are observed for $v\to v_f$ when the apparent 
horizon tends towards event horizon. If $\delta v$ exceeds $10^{3}\delta u$ apart from the vicinity of $v=v_f$ 
we regard this region as the temporary horizon. We assign the value of retarded time corresponding to the 
lowest value of $v$ with $\delta v\geqslant 10^{3}\delta u$ to the $u$-location of the temporary horizon.
\par
Resuming our discussion of spacetime structures, in both cases of negative $\alpha$ 
the location of 
the event horizon decreases smoothly in the retarded time $u$ while $\tpe_k$ increases. The $u$-locations 
of the singularity origin and temporary horizon also decrease with an increasing $\tpe_k$ smoothly apart 
from a small range of $\tpe_k$ values, where the changes are sudden. Up to around $\tpe_k=0.065$ 
for $\alpha=-\sqrt{3}$ and $\tpe_k=0.07$ for $\alpha=-1$, there is only spacelike singularity within 
the event horizon and the structure corresponds to typical Schwarzschild spacetime (the temporary 
horizon is absent). For larger values of the dilaton field amplitude the temporary horizon appears. 
Its appearance is bounded with a drastic change of the singularity origin $u$-location towards 
smaller values of the retarded time.\\
In the region corresponding to small values of $\tpe_k$ the lines indicating changes in the event horizon 
and the singularity origin locations are practically parallel. On the other hand, in the region, where 
the temporary horizon is present, the changes in its location are parallel to the ones corresponding 
to the event horizon. At the same time the location of the singularity origin tends towards the lower 
values of $u$-coordinate more slowly than both horizons.\\
Considering the comparison between the case of $\alpha$ equal to $-\sqrt{3}$ and the point 
of reference (RN spacetime) it should be stated that for all values of $\tpe_k$ the event 
horizon is situated along smaller value of $u$-coordinate when the dilaton field is present. On the other hand, the 
singularity origin appears at larger retarded time than in the case of reference for the values of $\tpe_k$ 
not exceeding $0.0525$. When the value of $\alpha=-1$ is taken into account, the event horizon is 
located at larger values of $u$ than in RN spacetime for the values of $\tpe_k$ smaller than $0.035$. The 
singularity origin corresponds to larger $u$ up to the point of the drastic decrease, that is up to $\tpe_k=0.0675$.

\subsubsection{Null dilatonic coupling constant}
\begin{figure}[p]
\includegraphics[scale=0.45]{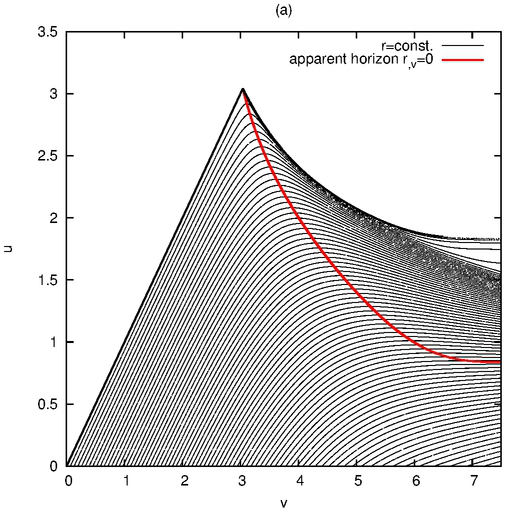}\includegraphics[scale=0.45]{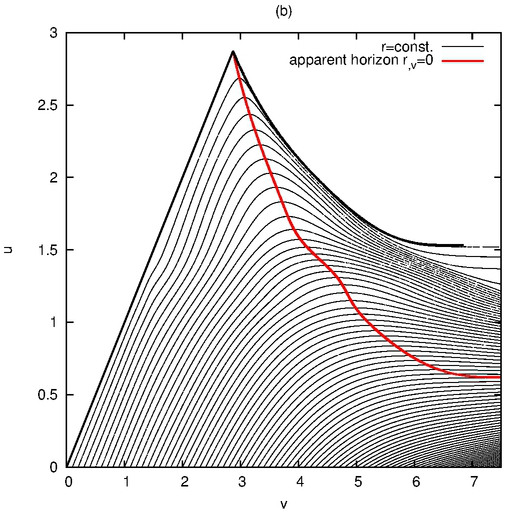}
\includegraphics[scale=0.45]{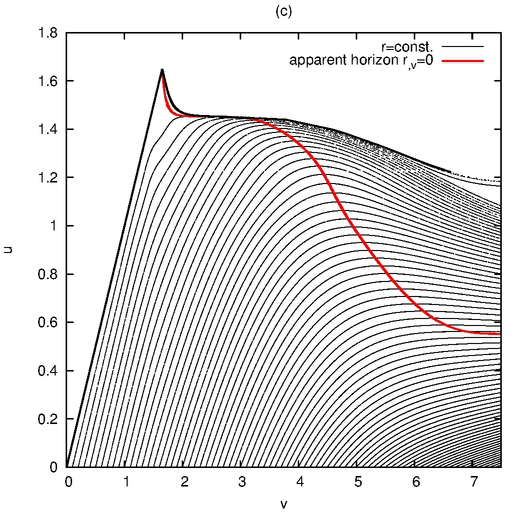}\includegraphics[scale=0.45]{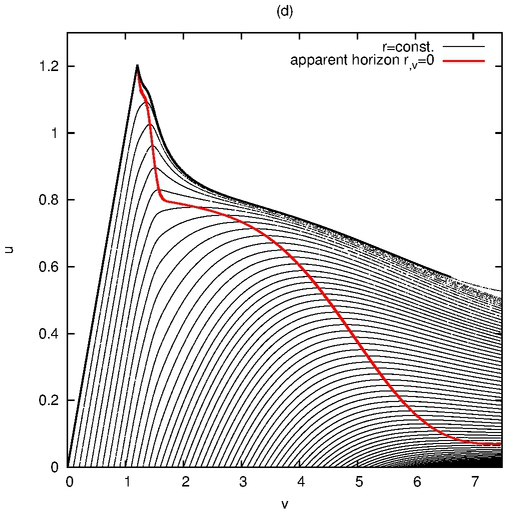}
\caption{Lines of constant $r$ in the $(vu)$-plane for the evolution with 
electric and dilatonic coupling constants equal to $e=0.5$ and $\alpha=0$, respectively. 
Family parameter for electrically charged scalar field is $\tpe_s=0.6$, while for dilaton field 
are chosen as 
(a) $\tpe_k=0.001$, (b) $\tpe_k=0.06$, (c) $\tpe_k=0.07$ and (d) $\tpe_k=0.13$.}
\label{fig04}
\end{figure}

Quite different structures of spacetimes are obtained as results of the dynamical collapse of the charged scalar field,
when the dilaton field is not coupled to $U(1)$-gauge Maxwell field, i.e., when $\alpha = 0$.
The amplitudes 
of the dilaton field in this case were set as equal to $0.001$, $0.06$, $0.07$ and $0.13$. 
The Penrose diagrams representing the structures of the corresponding spacetimes are shown in Fig.\ref{fig04}.
In comparison with the cases of non-zero dilaton coupling, the most significant difference
is that for all 
values of the parameter $\tpe_k$, the Cauchy horizon remains in the final state of the evolutions under consideration. 
The collapse finally leads to spacetimes of the RN-type, i.e., there are two horizons in spacetime, namely 
the event and the Cauchy ones as well as the central spacelike singularity within them.\\
For small values of the parameter $\tpe_k$ 
one gets RN structure of the spacetime.
The apparent horizon tends smoothly towards smaller values of $u$, approaching the event 
horizon at $v\rightarrow\infty$. Simultaneously, 
a collection of lines $r=const.$ settling along constant $u$-coordinate
appears indicating the existence of the Cauchy horizon in spacetime.
For larger values of the dilaton field amplitude 
its influence begins to be visible. Just as in both previously described cases, the bends in $r=const.$ 
lines and in the apparent horizon are visible for $u$ around $1.5$ (this time for $\tpe_k=0.06$, Fig.\ref{fig04}b).
But again, the dilaton field strength is still too small to cause any significant changes in the spacetime structure. 
Only for $\tpe_k$ exceeding $0.07$ its influence is of a greater importance. Similarly to the cases of the negative 
dilatonic coupling constant, the collapse runs in two stages. The first dynamical part of the considered evolution 
results in an appearance of a single temporary horizon. 
Next, the second part of the dynamical collapse emerges. The horizon changes its position 
to the smaller value of $u=const.$ and simultaneously the tendency of lines $r=const.$ for settling 
along $u=const.$ appears. As was already mentioned, the final spacetime contains 
two horizons, namely the event and Cauchy ones.
For spacetimes containing the temporary horizon the singularity origin appears for the considerably smaller 
values of the retarded time $u$, as may be seen in Fig.\ref{fig04}. In this case, it seems that 
thanks to the occurrence of the Cauchy horizon we are able 
to determine the meaning of each of the dynamical stages. The first one seems to be a collapse of the dilaton field 
and the second one may be identified with the electrically 
charged scalar field falling onto the newly born black hole.
\par
The results of a set of computations conducted for a vast range of $\tpe_k$ values for $\alpha=0$ 
are presented in Fig.\ref{fig06}c. The qualitative analysis of the changes of particular $u$-locations 
of the temporary and event horizons as well as the singularity origin locations is identical to the 
one performed above for the cases of $\alpha$ equal to $-\sqrt{3}$ and $-1$. Hence, we will not repeat it. The 
only fact that should be noted is that the sudden change in the location of the singularity origin accompanied 
by an appearance of the temporary horizon takes place around $\tpe_k=0.07$ when $\alpha=0$.\\
When comparing the results shown in Fig.\ref{fig06}c with the point of reference it turns out that both 
the event horizon and the singularity origin locations appear at smaller values of $u$-coordinate than in 
the case of reference. 
Moreover, it seems that in the considered case without dilaton coupling the reference spacetime is the 
limit while $\tpe_k\rightarrow 0$. It means that in order to get the RN spacetime it is not sufficient 
to make the initial dilaton field amplitude smaller disregarding the value of dilatonic coupling constant. In 
fact the coupling between the dilaton and Maxwell fields should also be excluded.

\subsubsection{Positive dilatonic coupling constant}
\begin{figure}[p]
\includegraphics[scale=0.45]{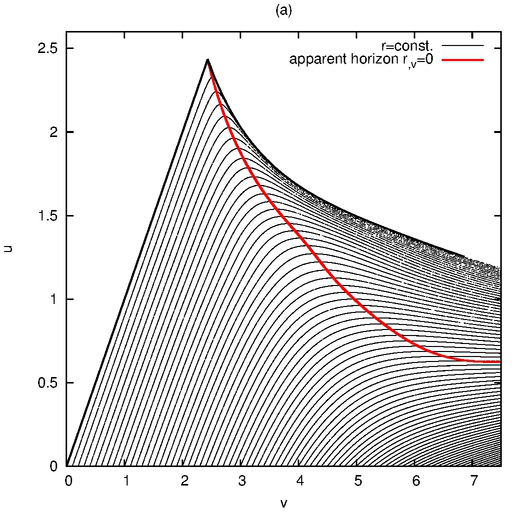}\includegraphics[scale=0.45]{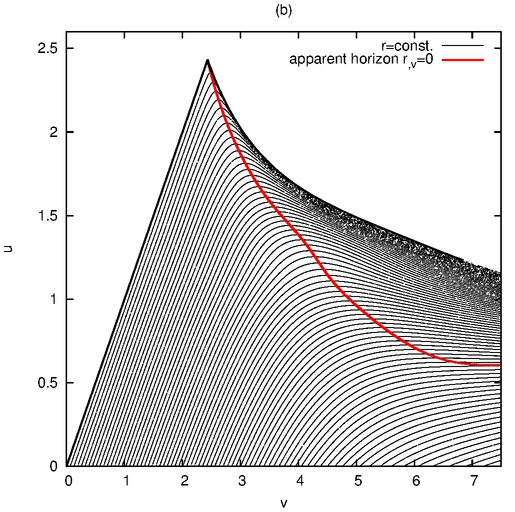}
\includegraphics[scale=0.45]{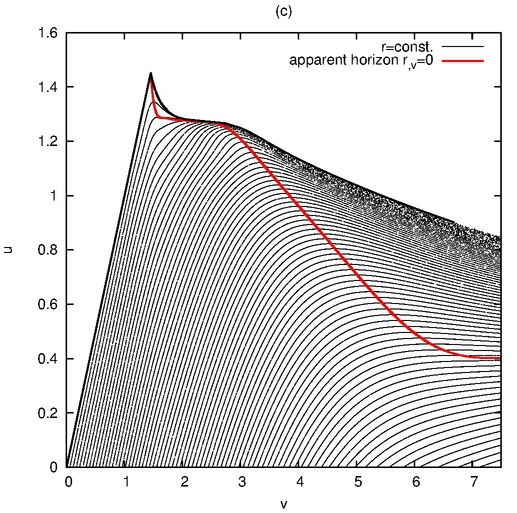}\includegraphics[scale=0.45]{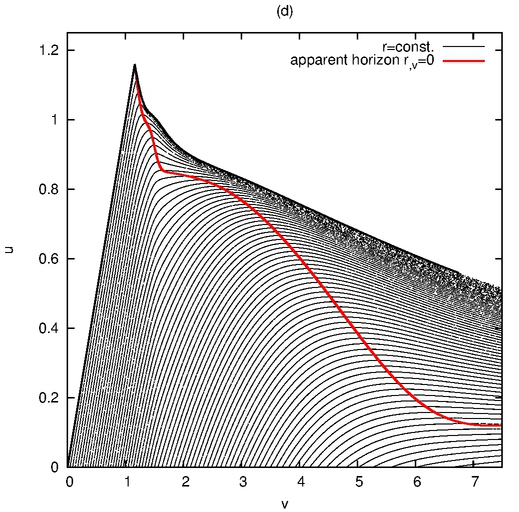}
\caption{Lines of constant $r$ in the $(vu)$-plane for the dynamical collapse 
with electric and dilatonic coupling constants equal to $e=0.5$ and $\alpha=+1$, respectively. 
Family parameter for electrically charged scalar field is $\tpe_s=0.6$, while for dilaton field are set (a) $\tpe_k=0.015$, (b) $\tpe_k=0.025$, (c) $\tpe_k=0.075$ and (d) $\tpe_k=0.12$.}
\label{fig05}
\end{figure}

\begin{figure}[p]
\includegraphics[scale=0.45]{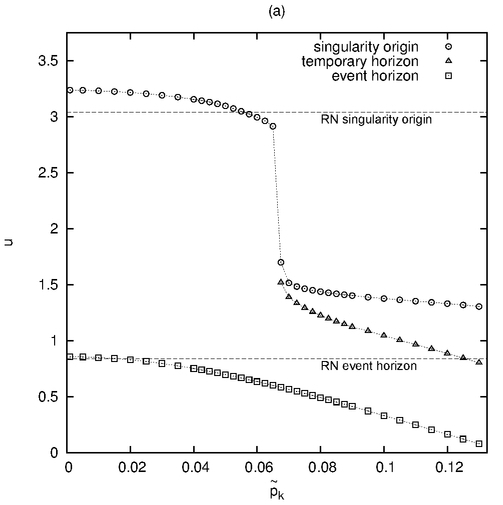}\includegraphics[scale=0.45]{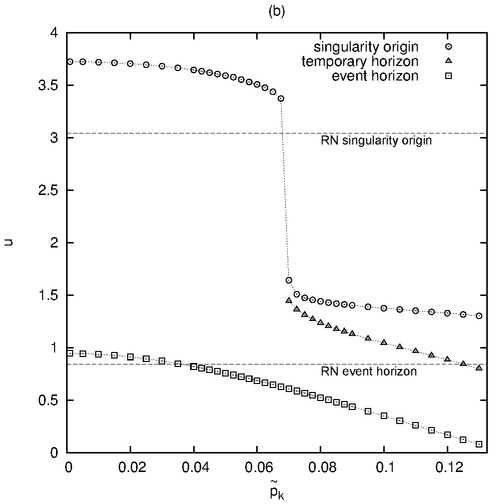}
\includegraphics[scale=0.45]{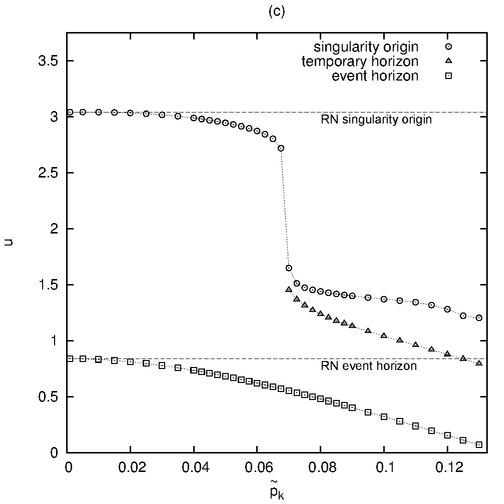}\includegraphics[scale=0.45]{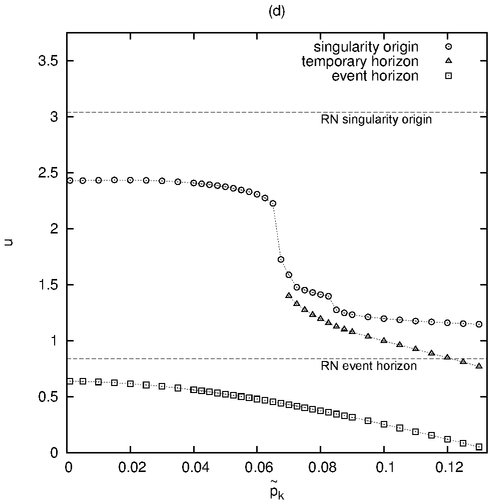}
\caption{The $u$-locations of temporary and final black hole's event horizons as well as the point, where the line $r=0$ becomes singular as functions of the altering amplitude of dilaton field $\tpe_k$ for the examined collapse. The free family parameter for electrically charged scalar field was set $\tpe_s=0.6$ and electric coupling constant was equal to $e=0.5$. The values of dilatonic coupling constant were set as equal to (a) $\alpha=-\sqrt{3}$, (b) $\alpha=-1$, (c) $\alpha=0$ and (d) $\alpha=+1$. The values of $u$-coordinate corresponding to locations of the event horizon and the singularity origin in the case of reference (RN spacetime) were depicted on each diagram.}
\label{fig06}
\end{figure}

We complete the second stage of the results interpretation by discussing the structures
of spacetimes emerging from the dynamical collapse when the coupling constant $\alpha$
is equal to $+1$. The results of the numerical computations are presented in Fig.\ref{fig05}.
In the considered case we choose the dilaton field amplitudes to be equal to 
$0.015$, $0.025$, $0.075$ and $0.12$. 
In general, the results are qualitatively similar to the ones described above, for $\alpha = -\sqrt{3}$ 
and $\alpha = -1$. The obtained spacetimes are Schwarzschild-type with one horizon surrounding central 
singularity. For small values of the initial dilaton field amplitude the typical structure is obtained, while a 
visible tendency towards black hole formation is noticeable for $\tpe_k$ around $0.03$ and two dynamical 
stages of the collapse appear for its larger values exceeding $0.07$.
\par
The locations of the event and temporary horizons and the singularity origin for a set of the dilaton field
initial amplitudes for $\alpha=+1$ are shown in Fig.\ref{fig06}d. Again, they are qualitatively similar to 
the
all previously discussed cases. The only new feature is that apart from the sudden decrease of the $u$-location 
of the singularity origin around $\tpe_k=0.07$ there is another one around $\tpe_k=0.085$. Yet it is not connected 
with any sudden changes in the temporary or event horizon locations.\\
The comparison to the case of reference reveals that the locations both of the event horizon and the singularity 
origin for all the $\tpe_k$ values correspond to the considerably smaller values of the retarded time than in RN spacetime.

\subsection{Properties of the dynamically formed black holes}
\begin{figure}[p]
\includegraphics[scale=0.45]{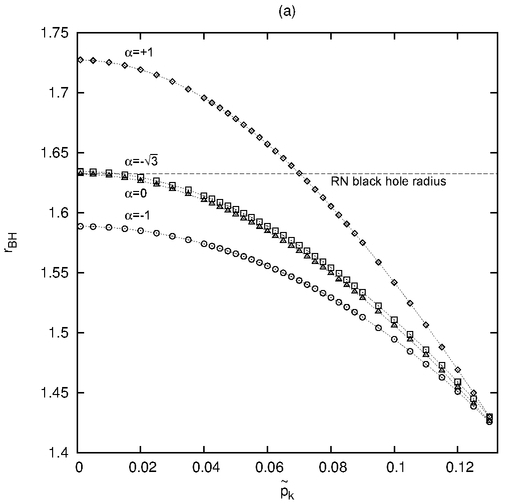}
\includegraphics[scale=0.45]{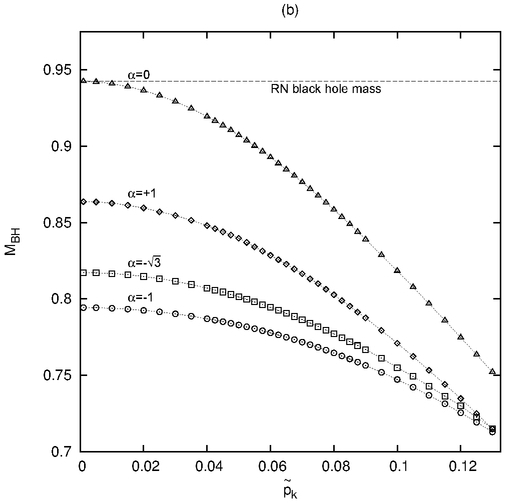}
\includegraphics[scale=0.45]{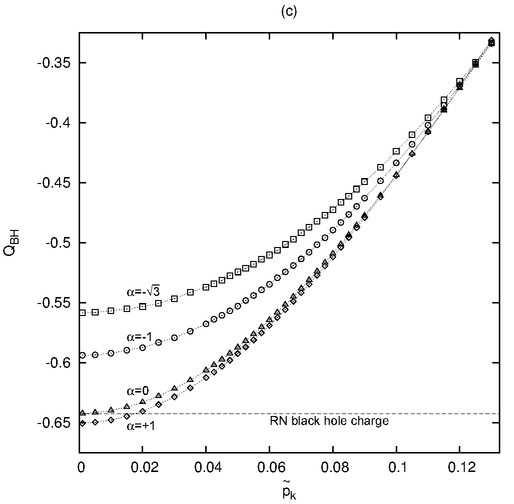}
\caption{The values of (a) radius, (b) mass and (c) charge of a black hole as functions of the altering amplitude of dilaton field $\tpe_k$ for dilatonic coupling constant $\alpha$ equal to $-\sqrt{3}$, $-1$, $0$ and $+1$. The free family parameter for electrically charged scalar field and electric coupling constant have the same values as in Fig.\ref{fig06}. The values of radius, mass and charge of a black hole present in the case of reference (RN spacetime) were depicted on corresponding diagrams.}
\label{fig07}
\end{figure}

Now, we shall focus our attention on the features of black holes present in spacetimes under consideration. We 
will comment on the values of black holes radii, masses and charges as functions of initial amplitudes of 
dilaton field $\tpe_k$ for the considered values of the dilatonic coupling constant $\alpha$. These plots are 
presented in Fig.\ref{fig07}. It should be stressed that we used two distinct methods of calculating the 
masses of black holes. For $\alpha=-\sqrt{3},\pm 1$ we used the expression (\ref{mhaw}), while for $\alpha=0$ 
the relation (\ref{mhaw-RN}) was used. Our choice was determined by the structures of emerging 
spacetimes (Schwarzschild-type in the former case and RN-type in the latter one).\\
In Fig.\ref{fig07}a we presented the values of the radius of a black hole for a wide range 
of the dilaton field initial amplitudes for all analyzed dilatonic coupling constants. It turns 
out that for all $\alpha$ the radius decreases as the initial amplitude of the dilaton field increases. Moreover, for 
increasing $\tpe_k$ it tends towards the same value for all $\alpha$. What may be also 
seen from Fig.\ref{fig07}a is that for a particular value of $\tpe_k$ the radii of 
black holes for different $\alpha$ vary as follows: 
\be
r_{BH,\alpha=-1}<r_{BH,\alpha=0}<r_{BH,\alpha=-\sqrt{3}}<r_{BH,\alpha=+1}.
\ee
The masses of black holes as functions of $\tpe_k$ for the considered values of dilatonic 
coupling constant are shown in Fig.\ref{fig07}b. For all values of $\alpha$ the masses decrease as 
$\tpe_k$ increases. We think that such behaviour is connected with the fact that the radius of a black 
hole decreases since black hole Hawking mass is the mass included in a sphere of radius corresponding 
to the event horizon. The masses tend towards the same value as $\tpe_k$ increases for $\alpha$ not 
equal to zero. For a particular value of $\tpe_k$ the masses of black holes for different 
$\alpha$ imply
\be
M_{BH,\alpha=-1}<M_{BH,\alpha=-\sqrt{3}}<M_{BH,\alpha=+1}<M_{BH,\alpha=0}.
\ee
In Fig.\ref{fig07}c we depicted the black hole charge as a function of the dilaton field initial 
amplitude for all $\alpha$ under consideration. All the charges are negative, but the sign is 
of no significance here, so we will comment on their absolute values. The charges of black holes 
decrease tending towards the same value with an increasing $\tpe_k$. It is also connected with 
the fact that at the same time the radius is getting smaller. For a particular value of $\tpe_k$ 
the absolute values of black hole charges for different $\alpha$ vary in the following manner: 
\be
Q_{BH,\alpha=-\sqrt{3}}<Q_{BH,\alpha=-1}<Q_{BH,\alpha=0}<Q_{BH,\alpha=+1}.
\ee
The comparison of the discussed features of black holes with the case of reference confirm the conclusion drawn 
from the analysis of spacetime structures performed above. It happens
that the RN spacetime is the limit of the case with $\alpha=0$ as may be inferred from Fig.\ref{fig07}. The radii, the 
masses and the charges of black holes resulting from the collapse without the dilatonic coupling tend towards the 
corresponding values determined for the reference RN spacetime as $\tpe_k\rightarrow 0$.
\par
\begin{figure}[p]
\includegraphics[scale=0.45]{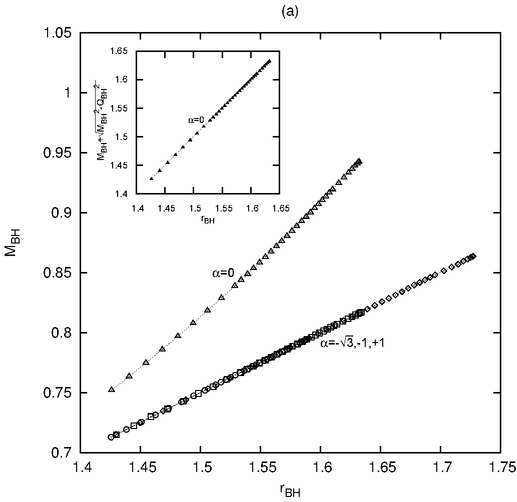}\\
\includegraphics[scale=0.45]{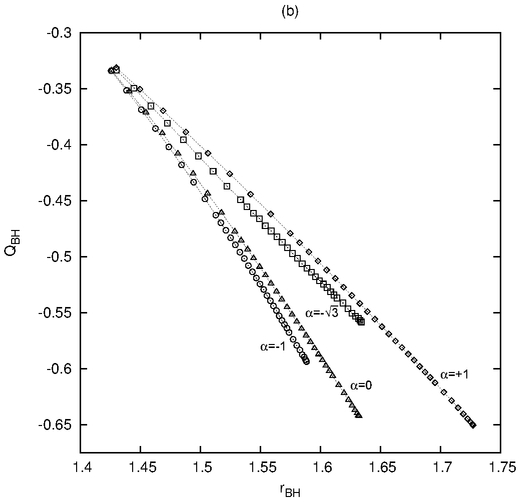}
\caption{The values of (a) mass and (b) charge of a black hole as functions of its radius for dilatonic coupling constant $\alpha$ equal to $-\sqrt{3}$, $-1$, $0$ and $+1$. The subfigure in (a) shows the relation between an expression $M_{BH}+\sqrt{M_{BH}^2-Q_{BH}^2}$ and black hole radius for $\alpha=0$. The free family parameter for electrically charged scalar field and electric coupling constant have the same values as in Fig.\ref{fig06}.}
\label{fig08}
\end{figure}

To complete the analysis of black hole features we will comment on the relations between the 
mass and charge of a black hole and its radius for all the values of $\alpha$ taken into account. The relations 
are shown in Fig.\ref{fig08}. The most straightforward conclusion is that both the mass and the absolute value 
of the charge increase as the radius increases. Not only the black hole mass increases linearly 
with radius, but also the mass dependence on radius is identical in all the cases of $\alpha\neq 0$ 
(Fig.\ref{fig08}a). The slope of the line $M_{BH}$ vs. $r_{BH}$ in these cases is equal to $0.5$. 
This result is consistent with the fact that the relation $r=2M$ is fulfilled along 
the event horizon of a Schwarzschild black hole.\\
In case of $\alpha=0$ the relation between radius and mass is not linear. It is a result of 
a different spacetime structure, namely RN-type in this case. In contrast to the 
Schwarzschild-like black holes, the relation between mass and radius is modified by the 
presence of the electric charge. In fact, the expression $r=M+\sqrt{M^2-Q^2}$ holds along the 
event horizon of RN black hole. As may be inferred from the sub-figure in Fig.\ref{fig08}a, this relation is 
fulfilled along the event horizon for black holes obtained during the collapse considered by us for $\alpha=0$.\\
The changes of the absolute value of a black hole charge with its radius are linear only in two 
cases of $\alpha$ equal to $-\sqrt{3}$ and $+1$ (Fig.\ref{fig08}b). The slopes of the 
lines $Q_{BH}$ vs. $r_{BH}$ are in these cases similar and are equal to $1.1$ and $1.08$, respectively. In 
the remaining cases the absolute value of the charge increases more quickly than the radius.
\par
Let us give some remarks concerning the physical interpretation of the considered numerical results. 
Because of this fact we shall try to 
enlighten some analogies between obtained results and the 
analytical form of the static black hole solution in dilaton gravity with arbitrary coupling 
constant $\alpha$ \cite{dil}.
\par 
First of all one can see that the exponential form of the coupling between dilaton and $U(1)$-gauge field
plays the crucial role in the dynamical collapse. Dilaton field effectively suppresses the emergence 
of the black hole structure resembling 
RN black hole. On the contrary, the collapse goes like in the Schwarzschild case. This conclusion 
finds its justification
in the analytic form of the static dilatonic black hole line element
which mostly resembles Schwarzschild black hole except the modification in
$S^2$-sphere radius, which depends on the charge and mass of the black hole.

In analytic treatment of the static dilaton black hole with arbitrary coupling constant 
one obtains the inner horizon 
which is invisible in our numerical studies of the dynamical collapse. However the tendencies 
of the behaviour of black hole characteristics such as mass, charge and the event horizon 
radius are similar. Namely, in analytical treatment one obtains
that the resulting mass of black hole is a sum of $r_{+}$ and $r_{-}$ multiplied by a constant coefficient, 
where $r_{+}$ and $r_{-}$ denote the radius of an outer and inner horizon, respectively.
If dilaton field increases the value of the inner horizon  $r_{-}$ decreases. Then, we achieve a general tendency that the 
bigger value of the dilaton field is considered the smaller value of mass one gets. The same conclusion was confirmed by 
the studies of the dynamical case.\\
On the other hand, in analytical computations the charge of the dilaton black hole is
a product of inner and outer horizon radii divided by a constant coefficient. 
Just, when dilaton field grows the inner 
horizon decreases and this in turn implies that the charge is smaller.
The same conclusion was drawn from numerical studies of the dynamical collapse in dilaton gravity.\\
Finally, when one considers the area of the black hole event horizon, analytical considerations reveal that the square of
$S^2$ sphere radius is a function of $r_{-}$. Then, one concludes that the the bigger value of 
dilaton field we take into 
account the smaller area of the event horizon one obtains. In numerical considerations this is also the case.

\section{Conclusions}
In our paper we have studied the influence of the dilaton field amplitude 
on the dynamical collapse of the charged scalar field.
The underlying system of equations was solved in the double-null coordinates. They enabled us to start with the regular 
initial spacetime at approximately null infinity, elaborate the formation of a black hole and extend our considerations to the 
singularity formed during the dynamical collapse process. The problem in question was formulated in the system of the
first order partial differential equations. The equations of motion and the numerical method of the problem
were elaborated in our previous article \cite{bor11}.
\par
After choosing the constant amplitude of the charged scalar field we conducted our studies for various 
values of the dilaton field amplitude. We also altered the value of the dilatonic coupling constant in 
the considered theory and checked the behaviour of the system for $\alpha$ equal to $-\sqrt{3}$, $-1$, $0$ and $+1$.
\par
It turns out that for all values of dilatonic coupling constant $\alpha\neq 0$ in the region 
of $v\rightarrow\infty$ the Schwarzschild black hole is obtained. Despite the presence of electric 
charge the spacetime contains one horizon surrounding central spacelike singularity. Moreover, the 
relation $r=2M$ is fulfilled along the event horizon. For $\alpha=0$ and $v\rightarrow\infty$ the 
structure of RN-type spacetime results from the evolution, i.e., there are two horizons and a 
central singularity in spacetime. The expression $r=M+\sqrt{M^2-Q^2}$ holds along the event horizon 
of black holes obtained in this case. These dynamically obtained structures are in 
agreement with the analytical solutions provided for the non-dynamical cases \cite{dil}.
\par
For small values of the initial dilaton field amplitude the course of a collapse is typical. The lines 
indicating the locations of the apparent horizon
and the line $r=0$ are smooth. For larger values of $\tpe_k$ the evolution runs in 
two stages. The temporary horizon appears, which means that the apparent horizon settles along $u=const.$ 
temporarily and then changes its $u$-location towards smaller values of retarded time as $v\rightarrow\infty$. We 
think that such behaviour is connected with the fact that one of the fields collapses earlier than the other 
(in fact dilaton field before the electrically charged scalar field). This statement is hard to justify 
in case of $\alpha\neq 0$, but when $\alpha=0$ the appearance of the Cauchy horizon
during the second stage of a collapse provides a proof of our proposition.
\par
The characteristics of black holes obviously change when the value of an initial dilaton field 
amplitude changes. For all analyzed values of $\alpha$ the radii, masses and absolute values of 
charges of black holes decrease while $\tpe_k$ increases.
Such behaviour of the properties of a black hole is typical of the analytical dilaton black hole. For this 
reason we conclude that the dilaton field has superior status in comparison with the electrically charged 
scalar field during the dynamical collapse under consideration.
\par
Both spacetime structures and the features of black holes present in these spacetimes indicate 
that the RN spacetime (pure collapse of an electrically charged scalar field, without dilaton field) 
is the limit of the evolutions when $\tpe_k\rightarrow 0$ only when $\alpha=0$. For non-zero dilatonic 
coupling constant the spacetime structures do not tend towards RN spacetime when $\tpe_k$ approaches zero.

\begin{acknowledgements}
AN was supported by Human Capital Programme of European Social Fund sponsored by European Union.
MR was partially supported by the grant of National Science Centre 2011/01/B/ST2/00488.
We are grateful David Langlois for valuable comments and discussions.
\end{acknowledgements}





\end{document}